\documentclass[aps,prc,twocolumn,groupedaddress,showpacs,showkeys]{revtex4-1}

\usepackage{natbib}

\usepackage{amssymb}

\usepackage{amsmath}

\usepackage{graphicx}

\usepackage{bm}

\begin{document}

\title{pp forward elastic scattering amplitudes at 7 and 8 TeV}

\author{A. K. Kohara}
\email{kendi@if.ufrj.br}
\author{E. Ferreira}
\email{erasmo@if.ufrj.br}
\author{M. Rangel}
\email{mrangel@if.ufrj.br}
\affiliation{Instituto de F\'{\i}sica, Universidade Federal do Rio de Janeiro, 21941-972, Rio de Janeiro, Brazil}

\author{T. Kodama}
\email{tkodama@if.ufrj.br}
\affiliation{Instituto de F\'{\i}sica, Universidade Federal do Rio de Janeiro, 21941-972, Rio de Janeiro, Brazil}
\affiliation{Instituto de F\'{\i}sica, Universidade Federal Fluminense, Niter\'oi, Brazil}

\date{\today}

\begin{abstract}
We analyse the recent LHC data at 7 and 8 TeV for pp elastic scattering with special attention for the structure of the real part, which  is shown to be crucial to describe the differential cross section in the forward region. We determine accurately the position of the zero of  the real amplitude, which corresponds to the zero of a theorem by Andr\'e Martin. 

\end{abstract}




\maketitle

\section{Introduction}

The elastic amplitude $T(s,t)$ is a function of only two kinematical variables, controlled by principles of analyticity and unitarity, but no fundamental solution is known for its form, and representations of the differential cross section are given in terms of models, designed and applied for restricted ranges of $s$ and $t$. It is expected that at high energies the $s$ dependence becomes relatively simple, but the enormous gaps and uncertainties in the data from CERN/ISR, Fermilab and CERN/LHC do not help in tracing the $s$ dependence with reliability. On the other hand, for a given $s$, the angular dependence has not been measured with uniformity in the full $t$-range, and the necessary disentanglement of the real and imaginary parts of the amplitude is a hard task, with unavoidable indetermination \cite{LHC7TeV,LHC8TeV}. The forward $t$ range has been measured more often. Recently Totem and Atlas  groups at LHC measured $d\sigma/dt$ in forward $t$ ranges at $\sqrt{s}$ = 7 and 8 TeV \cite{T7, A7, T8, A8}. These data (Table \ref{datasets}) offer an opportunity to study in detail several aspects of the very forward region, such as the magnitudes of the real and imaginary amplitudes, the position of the zero of the real part and the first derivatives of the amplitudes with respect to the variable $t$ (slopes). 
   The Coulomb-nuclear interference  depends on the proton electromagnetic structure, and the relative phase requires specific assumptions for the forms of the nuclear  amplitudes as described in detail in our recent work \cite{us}. 
   

In the present work we propose independent parametrizations for the real and imaginary nuclear parts, writing
\begin{equation}
    T_R^N(t)= [1/(4 \sqrt{\pi} \left(\hbar c\right)^2)]~ \sigma (\rho -\mu_R t) ~ e^{B_R t/2}  ~,
\label{real_TR}
\end{equation}
and 
   \begin{equation}
T_I^N (t)= [1/(4 \sqrt{\pi} \left(\hbar c\right)^2)] ~ \sigma ( 1 -\mu_I t )~ e^{B_I t/2} ~ .
\label{imag_TI}
\end{equation}
The parameter $\sigma$ is the total cross section, $\rho$ is the ratio of the real and imaginary parts at $|t|=0$, $B_R$ and $B_I$ are the local slopes of the amplitudes and the parameters $\mu_R$ and $\mu_I$ account for the existence of zeros in the amplitudes. The zero in the real part is crucial to explain the $t$ dependence of $d\sigma/dt$ for small $|t|$.



We remark that parameters are determined fitting data in limited $|t|$ ranges, at finite 
distance from the origin, so that the values obtained  depend on the 
analytical forms  (\ref{real_TR},\ref{imag_TI})
of the amplitudes. 
In particular, the slope parameter usually written in   form 
  $  d\sigma/dt= \sigma^2 (\rho^2+1) \exp(Bt)$ 
does not agree with the expression for the differential cross section as
sum of  two independent squared  magnitudes, each with its own slope. The assumption 
that $B_R$ 
and $B_I$ are equal is not justified. The  average  quantity  $B$ alone gives  rough and 
unsatisfactory information.  
The importance of the different slopes in the analysis of pp elastic scattering has 
been investigated in the framework of the so called dispersion relations for slopes \cite{EF2007}. 
It is important to note that also   the Coulomb-Nuclear 
phase $\phi(t)$ depends essentially on the form of the  nuclear amplitudes \cite{KL}.
The zero of the real part is  given  by  $|t_R|=-\rho/\mu_R$. We understand that this quantity is the zero predicted in the theorem by Andr\'e Martin \cite{Martin}.

Of course the parameters of the amplitudes are correlated, and in the present work we investigate the 
bounds of the correlations. We attempt  to identify  values 
of parameters that may be considered as common representatives for different  
measurements. We show  that the  differences between the two experimental collaborations 
may  be restricted  to  quantities characterizing normalization.  
 The question of normalization is essential, and our inputs are  the values
of $d\sigma/dt$ given in the experimental papers  \cite{T7,A7,T8,A8}. 

The extraction of forward parameters in pp scattering  has difficulties due 
to the small value of the $\rho$ parameter, and consequently  has suffered in many 
analyses  from neglect of the properties of the real part.   
In our view the values of $\sigma$, $\rho$, $B$ appearing in universal databases
\cite{PDG,COMPETE} 
as if they were direct experimental measurements should give room for 
critically controlled  phenomenological determinations. 
A proper consideration for the properties of the complex amplitude is necessary. 
We observe 
that the properties that $B_R \neq B_I$ and of the presence of zeros are common to 
several models  \cite{LHC7TeV,LHC8TeV,Models}. The determination 
of the amplitudes for all $|t|$ in several models is coherent. 

We observe that the polynomial factors written in the exponent in some 
parameterizations of data \cite{T8}
correspond to  the  linear and quadratic  factors mentioned 
above, if the assumption is made that they are much smaller than 1  and 
can be converted into exponentials. 
However  this substitution is not convenient, because it does not show explicitly the 
essential zeros, and it also gives unsatisfactory parameterization that cannot 
be extended even to nearby $|t|$ values.

 We thus have the framework necessary   for the analysis of the data,
    with clear identification of the role of the free  parameters.
  The    quantities to be determined for each dataset  are $\sigma$, $\rho$, 
 $B_I$, $B_R$, $\mu_I$, $\mu_R$.

\section{Data analysis  at $\sqrt{s}=$ 7 and 8 $\rm {TeV}$}

The analysed datasets and  their $t$ ranges   are listed   in Table \ref{datasets}, 
where  T7, T8, A7, A8  specify  Totem (T) and Atlas (A) 
Collaborations and center-of-mass energies 7 and 8 TeV. 
In the measured ranges the Coulomb effects play 
important role and the relative 
Coulomb phase is properly taken into  account \cite{us}.

In order to identify  values for parameters valid for all measurements,  we study four different conditions  in the fits: 
   I) all six  parameters  are  free  ;  
 II) fixing $\rho$ at 0.14, as suggested by dispersion relations;  
 III) fixing $\mu_I$  from  the expected positions of   imaginary zero \cite{LHC7TeV, LHC8TeV} and dip in $d\sigma/dt$; 
    IV) fixing simultaneously $\rho$ and $\mu_I$ at the above values. A complete table with the  results can be found in ref. \cite{us}. In the present work we show the values obtained with Condition IV) in Table \ref{Table:FINAL}. The first reason for this choice is the existence of a correlation between the parameter $\mu_R$ and the parameter $\mu_I$ of the imaginary amplitude, and since the dip structure is presented in pp elastic scattering for larger t values, and the position of the dip is intimately related with the parameter $\mu_I$, the determination of $\mu_I$ constrains the value of $\mu_R$. The second reason is justified because the parameter $\mu_R$ together with $\rho$ construct the position of the zero of Martin, which is suggested to have an asymptotic form $|t_R|\sim 1/\log^2 s$ \cite{LHC8TeV}. As mentioned in \cite{us} the data sets analysed (A7, A8, T7 and T8) do not cover the large $|t|$ range where the dip structure is presented and also, the values of $\rho$ are sensitive to the Coulomb phase, which depends on the structure of the nuclear amplitude, and is still an open question. In order to contour these difficulties we fix both $\rho$ and $\mu_I$  at their {\it expected}  values  and we obtain good modeling  for all measurements, except for the total cross sections, that distinguish Atlas from Totem.  
    
\begin{figure*}
  \includegraphics[width=7.8cm]{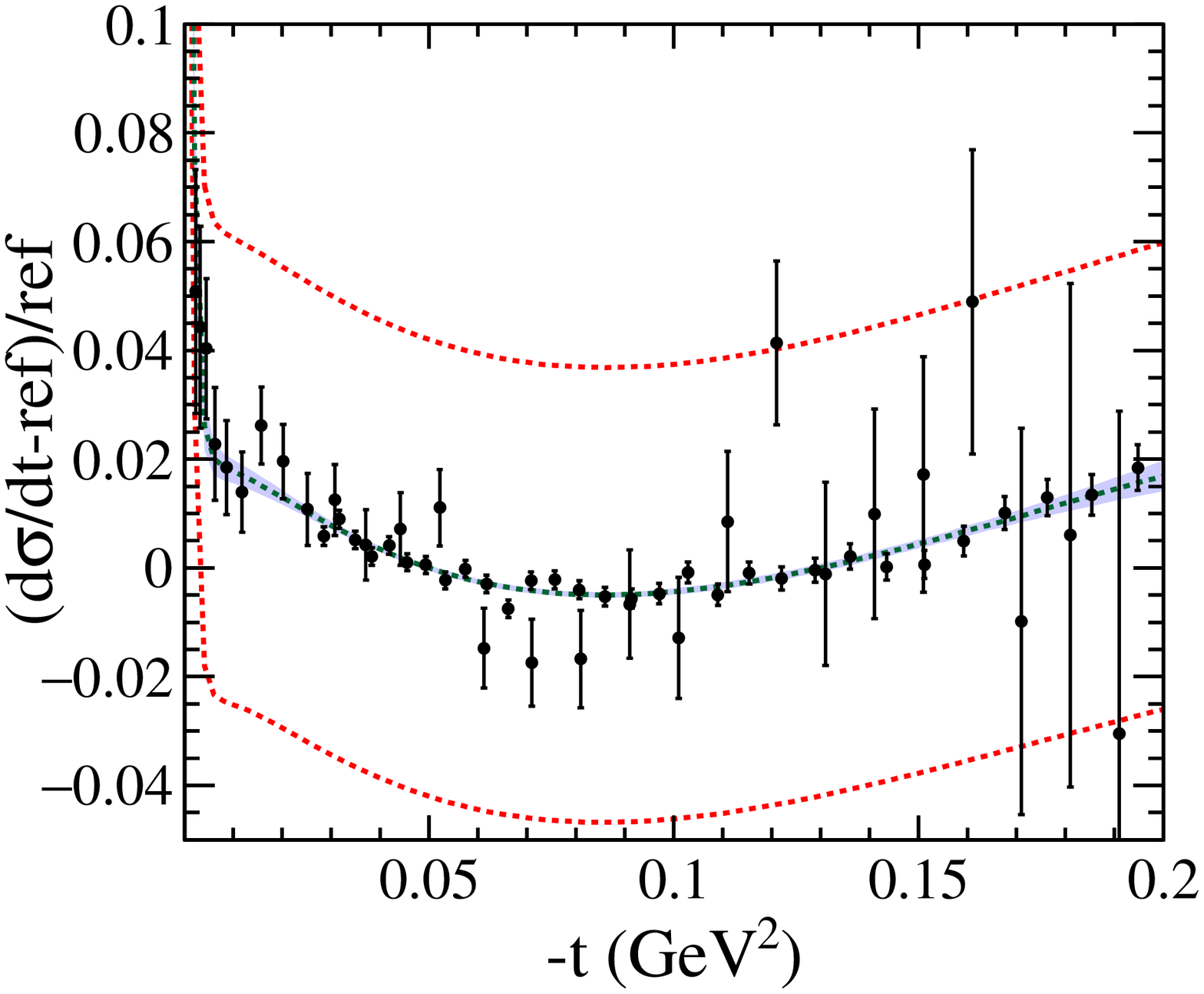} 
  \includegraphics[width=8.0cm]{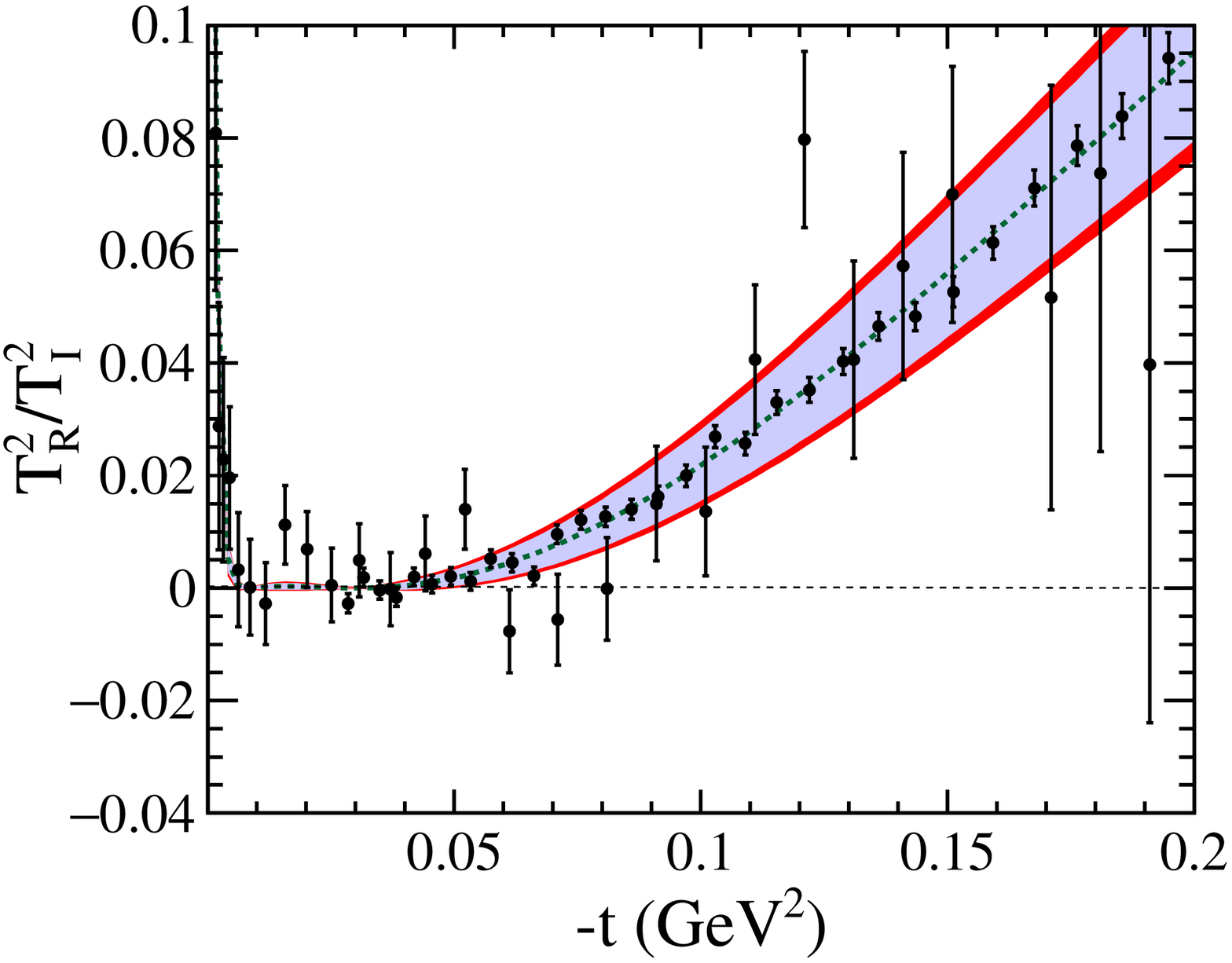}  
\caption{The left  plot  shows the non-exponential behaviour of the differential 
cross section for T8. The figure is obtained subtracting from the best fit  of the 
differential cross section a reference function which is $d\sigma/dt$ written 
 with a pure exponential form $ref = A\exp(B t)$  and dividing the subtraction by this reference function. 
The dashed lines show the normalization error band
in $d\sigma/dt$, that is quite large. The  plot in the RHS shows the ratio $T_R^2/T_I^2$ 
which exhibits information of a non-exponential behaviour with advantages compared with 
the first plot, since $\sigma$ is cancelled, and with it most of the normalization
systematic error.}
\label{displacement} 
\end{figure*}

      The regularity on the values of $\mu_R$ is remarkable, and the position of the zero 
   is stable in all measurements with  $|t_R|\simeq 0.037$ GeV$^{2}$ within the statistical errors. The position of the zero, together with the magnitude 
of $B_R$ determines the structure of the real amplitude. The zero of the real amplitude is responsible for the structure shown in  Fig. \ref{displacement}, where the differential cross section was subtracted by a pure exponential form called $ref = A\exp(B t)$ and divided by this quantity. Roughly speaking the $ref$ function has the similar structure and the same magnitude of the imaginary amplitude in the forward region, which means that the structure of a valley shown in l.h.s of Fig.\ref{displacement} is due to the structure of the real part. The r.h.s of Fig. \ref{displacement} is an alternative quantity that instead of working with the $ref$ function we have the squared of the real and imaginary amplitudes. The advantage of this language is that the errors due to the normalization of the cross section are suppressed leading to a much narrower error band. It is also interesting to observe that on the r.h.s of Fig.\ref{displacement}  at $|t|$ near 0.01 GeV$^{-2}$ the quantity $T_R^2/T_I^2$ has a zero due to the interference of the real and Coulomb amplitude, since for pp scattering the Coulomb amplitude is negative while the real nuclear is positive near the origin.  

The zero of the imaginary  part  anticipates the dip in the differential cross section that occurs beyond the range of the available data under study. 

Our analysis indicates  that the real amplitude plays crucial role in the description of the  differential cross section in the forward region.
Interference with the Coulomb interaction is properly accounted for, 
and use is made of  information from  external sources, such as dispersion relations and predictions for the imaginary zero obtained in studies of full-$t$ behaviour of the differential cross section \cite{LHC7TeV,LHC8TeV}.

\begin{table*}
 \vspace{0.5cm}
  \footnotesize
\begin{tabular}{ c c c c c c c c }
\hline
\hline
      &             &               &   &  &  &      &      \\
\hline 
  $\sqrt{s}$  & dataset  & $\Delta {|t|}$  range &  N         & Ref. &  $\sigma$ & $B_I$& $ \rho$        \\
     (GeV)    &               & (GeV$^{2}$)          &  points    &      &   (mb)    & ( GeV$^{-2})$ &     \\
\hline 
  7           &   T7       &  0.0052-0.3709      & 87         & 1    & 98.6$\pm$2.2      &19.90$\pm$0.30     & 0.14 (fix)$^{\rm a}$  \\
\hline
   7          &   A7       &  0.0062-0.3636        & 40        & 2     & $ 95.4\pm  0.4 $  & $19.73\pm 0.14 $ & 0.14 (fix) $^{\rm b}$       \\
 \hline
   8           &   T8       & 0.0007-0.1948      & 60       & 3     &  $103.0 \pm 2.3   $  &  19.56 $\pm$ 0.13     & (0.12 $\pm$ 0.03) $^{\rm c}$     \\
\hline
   8          &   A8       &   0.0105-0.3635       & 39        & 4     & $96.1\pm 0.2 $    &  $19.74\pm 0.05$     &  0.136  (fix)$^{\rm d}$    \\
 \hline
 \end{tabular}
 \caption{ Values of parameters at $\sqrt{s}=$7 and 8 TeV determined by Totem and Atlas Collaborations at 
LHC \cite{T7,A7,T8,A8}. Values for $\rho^{\rm [a]}$ and $\rho^{\rm [b]}$   are taken from COMPETE Collaboration \cite{COMPETE};  $\rho^{\rm [c]}$  obtained by the authors with 
a forward SET-I and kept fixed in a complete  SET-II; $\rho^{\rm [d]}$ is taken from \cite{PDG}.}  
\label{datasets}
\end{table*}

\begin{table*}
 \vspace{0.5cm}
 \footnotesize
\begin{tabular}{c c   c c c c c c c}
\hline
\hline
           &     &  &   &          &  &   &         \\
\multicolumn{8}{c}{Fixed Quantities :   $\rho = 0.14 $ , $\mu_I = - 2.16~ {\rm GeV}^{-2}$ (8 TeV) \cite{LHC8TeV}, $\mu_I = -2.14~ {\rm GeV}^{-2}$ (7 TeV) \cite{LHC7TeV} }  \\
\hline 
\hline 
&   N &$\sigma$       &$ B_{\rm I}  $      &$ B_{\rm R}  $      & $\mu_{\rm R}  $      & $-t_R$  &  $\chi^2/$ndf\\
  &         & (mb)                 &$({\rm GeV}^{-2})$ &$({\rm GeV}^{-2})$ & $ ({\rm GeV}^{-2})$ & $({\rm GeV}^{2})$ &         \\
      \hline
       \hline
T8&   60   &102.40$\pm$0.15& 15.27$\pm$0.39      &21.15$\pm$0.39      & -3.69$\pm$0.15      &   0.038$\pm$0.002 & 69.2/56  \\
  \hline
A8&  39   &96.82$\pm$0.11 & 15.26$\pm$0.06      &21.65$\pm$0.24      & -3.69$\pm$0.12      &  0.038$\pm$0.001  & 30.0/35  \\
 \hline
T7&   87    &99.80$\pm$0.21 & 15.71$\pm$0.14      &24.26$\pm$0.47      & -4.24$\pm$0.31  & 0.033$\pm$0.002 & 95.1/83  \\
\hline
T7& 87+17  &99.44$\pm$0.14 & 15.44$\pm$0.07      &22.62$\pm$0.19      & -3.49$\pm$0.13     & 0.040$\pm$0.002  & 203.5/100  \\
\hline
A7& 40   &95.75$\pm$0.16 & 15.23$\pm$0.11      &21.86$\pm$0.44      & -3.99$\pm$0.22      & 0.035$\pm$0.002  & 27.3/36  \\
\hline
\hline
\end{tabular}
 \caption{Proposed values of parameters for the  four datasets. The T7 data are also shown with inclusion of points at 
higher $|t|$ ($0.005149 < |t| < 2.443$ GeV$^2$)  that are important for confirmation of the value of $\mu_I$  \cite{us}. }
\label{Table:FINAL}
\end{table*}

\section{Conclusions   }

In this work we study the properties of the amplitudes in  pp elastic scattering 
analysing experimental data at the LHC center-of-mass energies 7 and 8 TeV, 
  based on a model for the 
complex amplitude, with explicit real and imaginary parts, each containing an 
exponential slope and a linear factor to account for the existence of a zero. 
The zero of the real part, 
close to the origin, corresponds to Martin's Theorem, and  the zero of the imaginary 
part  anticipates the dip in the differential cross section that occurs beyond 
the range of the available data under study.

Our study shows  that the real amplitude plays crucial role in the description of the 
differential cross section in the forward region.
Interference with the Coulomb interaction is properly accounted for, 
and use is made of  information from  external sources, such as dispersion relations 
and predictions for the imaginary zero obtained in studies of full-t behaviour 
of the differential cross section \cite{LHC8TeV,LHC7TeV}. 
  We organize the analysis under four conditions,
according to the specifications of the parameters with values fixed in each case. 
Comparison is made of the results obtained for  the four experimental 
measurements. We obtain the results shown in Table~\ref{Table:FINAL} that we believe to be 
a good representation of the experimental  data of Table \ref{datasets}.


The   quantity  $\mu_R$   is related  with the scaling variable $\tau = t \log^2{s}$ 
introduced by J. Dias de Deus   \cite{Deus} connecting $s$ and $t$ dependences 
in the amplitudes  at high energies and small $|t|$.
 A. Martin \cite{Martin_Real}  uses the same idea of a scaling variable,  writing 
an equation for the real part $\rho(s,t)$ using crossing symmetric scattering 
amplitudes   of a complex $s$ variable,  valid in a forward range. 
 The proposed ratio  is 
\begin{equation}
\rho(s,t)\simeq \frac{\pi}{\log s}\Big(1+\frac{\tau(df(\tau)/d\tau)}{f(\tau)}\Big) ~ ,  
\label{real_martin}
\end{equation}
where $f(\tau)$ is a damping function, with the implicit existence of a real zero.
The form  of $f(\tau)$ determines the properties of the real zero  \cite{Dremin}. 
that is found in the analysis of the data. 
This may be a clue for the introduction of explicit crossing symmetry and   
analyticity in our phenomenological treatment of the data.

Other models \cite{Models} also deal with the position of the real zero, 
discussing different analytical forms for the amplitudes, and it would 
be interesting to investigate their predictions for the amplitudes in the 
forward range.

In non-perturbative QCD, in several instances, the proton  appears as 
a structure with expanding size as the energy increases \cite{sizes}, 
with varied mechanisms, as distribution of valence quarks 
in a cloud around a core, modifications in QCD vacuum in the region of the 
colliding particles, and so on. 
Together with the evolution of the proton hadronic size, its electromagnetic 
properties, as they appear in high energy collisions, may change also. 
A linear increase in $\log{s}$ is a usual assumption for the effective proton radius,  
and   the form factor parameter $\Lambda^2$ would  then be reduced by about 1/2, 
corresponding to increase of   about 40 \% in proton radius. In Appendix A of reference \cite{us} we calculate the 
interference phase with this example.

We expect that future data in pp elastic scattering at 14 TeV 
will have high quality covering a wide $t$ range 
to allow  determination of the properties  of the real and imaginary 
amplitudes in pp elastic scattering, including studies of 
the   amplitudes up to the perturbative tail for large $|t|$.  
Hopefully the experimental groups will receive  the necessary  support and 
encouragement for this effort.   \\


\begin{acknowledgements} 
The authors wish to thank XIV Hadron Physics organizers for this stimulating Workshop. This work is a part of the Brazilian project INCT-FNA Proc. No. 464898/2014-5. The authors wish to thank the Brazilian agencies CNPq, CAPES for financial support. 
\end{acknowledgements}



\begin{thebibliography}{99}



\bibitem{LHC7TeV}  A. Kendi Kohara, E. Ferreira, and T. Kodama, \emph{Eur.
Phys. J. C} {\bf 73}, 2326  (2013). 

\bibitem{LHC8TeV} A. K. Kohara, E. Ferreira, and T. Kodama, \emph{Eur.
Phys. J. C} {\bf 74},  3175 (2014) .



\bibitem{T7}  G. Antchev et al. (TOTEM Coll.), \emph{Eurphys. Lett.} {\bf 101}, 21002 (2013).


\bibitem{A7}  G. Aad et al. (ATLAS Collaboration), \emph{Nucl. Phys. B} {\bf 889}, 486  (2014).

\bibitem{T8} G. Antchev et al. (TOTEM Coll.), \emph{Eur. Phys. J. C.} {\bf 16}, 661 (2016) ; 
Nucl. Phys. B {\bf  899},  527 (2015).


\bibitem{A8}  G. Aad et al. (ATLAS Collaboration), \emph{Phys. Lett. B} {\bf 761} 158 (2016).


\bibitem{us} A. K. Kohara, E. Ferreira, T. Kodama, and M. Rangel,  \emph{Eur. Phys. J. C} {\bf 77}, 877 (2017)

\bibitem{EF2007} E. Ferreira, \emph{Int. Jour. Mod. Phys. E} {\bf 16}, 2893 (2007).


\bibitem {KL}V. Kundr\'at and M. Lokajicek, \emph{Phys. Lett. B }
\textbf{611}, 102 (2005); V.Kundr\'at, M.Lokajicek and I. Vrococ,
\emph{Phys. Lett. B } \textbf{656}, 182 (2007).



\bibitem{Martin} A. Martin, \emph{Phys. Lett. B} {\bf 404}, 137 (1997).


\bibitem{COMPETE} J. R. Cudell et al. (COMPETE Collaboration), \emph{Phys. Rev. Lett.} {\bf 89}, 
              201801 (2002)


\bibitem{PDG} C. Patrign {\em et al.} (Particle Data Group), \emph{Chinese Physics C} {\bf 40}, 100001 (2016).


\bibitem{Models} C. Bourrely, J. Soffer, and T. T. Wu, \emph{Nucl. Phys. B} {\bf 247}, 15
~(1984); \emph{Phys. Rev. Lett.} {\bf 54}, 757 ~(1985); \emph{Phys. Lett. B} {\bf 196}, 237
~(1987); A. K. Kohara, E. Ferreira, and T. Kodama, \emph{Phys.
Rev. D} {\bf 87}, 054024 (2013); V. A. Petrov, E. Predazzi and A. V. Prokudin , \emph{\ Eur.
Phys. J. C } \textbf{28}, 525 (2003);
  O. V. Selyugin , \emph{Phys. Rev. D} \textbf{60}, 074028
(1999); M. M. Islam, R. J. Luddy, and A. V. Prokudin, \emph{Mod. Phys. Lett. A} {\bf 18}, 743 (2003); 
M.M. Islam and R.J. Luddy, \emph{Acta Phys. Pol. B Proc. Sup.}, {\bf 8}  4 (2015).


\bibitem{Deus} J. D. Deus, \emph{Nuc. Phys. B} {\bf 59}, 231 (1973)  ; \emph{Phys. Lett. B} {\bf 718},  1571 (2013). 
\bibitem{Martin_Real} A. Martin, \emph{Lett. Nuovo Cim.} {\bf 7},  811 (1973) . 


\bibitem{Dremin} I. M. Dremin, arXiv:1204.1914 [hep-ph] . 


\bibitem{sizes}   J. Dias de Deus and P. Kroll, \emph{Nuovo Cimento A} {\bf 37}, 67
~(1977); \emph{Acta Phys. Pol. B} {\bf 9}, 157 ~(1978); \emph{J. Phys. G} {\bf 9}, L81
~(1983); P. Kroll, \emph{Z. Phys. C} {\bf 15}, 67 ~(1982); T. T. Chou and C.
N. Yang, \emph{Phys. Rev.} {\bf 170}, 1591 ~(1968); \emph{Phys. Rev. D} {\bf 19}, 3268
~(1979); \emph{Phys. Lett. B} {\bf 128}, 457 ~(1983); \emph{Phys. Lett. B} {\bf 244}, 113
~(1990) ; 
  B. Povh and J. H\"ufner, \emph{Phys. Rev. Lett.} {\bf 58}, 1612 ~(1987); \emph{Phys.
Lett. B} {\bf 215}, 722 ~(1988); \emph{Phys.
Lett. B}  {\bf 245}, 653 ~(1990); \emph{Phys. Rev. D} {\bf 46},
990 ~(1992); \emph{Z. Phys. C} {\bf 63}, 631 ~(1994); 
E. Ferreira and F. Pereira, \emph{Phys. Rev. D} {\bf 55}, 130 (1997); \emph{Phys. Rev. D} {\bf 56}, 179 (1997). 
  


\end{thebibliography}
\end{document}